\documentclass{mem}
\usepackage{natbib}\usepackage{txfonts}\usepackage{balance}
\usepackage{graphicx}
\usepackage[a4paper]{hyperref}
\idline{xx}{1}

\begin{document}
\def\teff{$T\rm_{eff }$}
\def\kms{$\mathrm {km s}^{-1}$}

\title{
Binary post-AGB stars and their Keplerian discs
}

\subtitle{}

\author{
H. Van Winckel\inst{1} \and
T. Lloyd Evans\inst{2} \and
M. Reyniers\inst{1} \and
P. Deroo\inst{1} \and
C. Gielen\inst{1}  
 }

\offprints{H. Van Winckel}

\institute{
Instituut voor Sterrenkunde, KULeuven, Celestijnenlaan 200B, 3001
Leuven (Heverlee), Belgium, \email{Hans.VanWinckel@ster.kuleuven.be}  \and
School of Physics and Astronomy, University of St. Andrews, North
Haugh, St. Andrews, Fife KY16 9SS, Scotland
}

\authorrunning{Van Winckel}

\titlerunning{Binary post-AGB stars}

\abstract{

In this contribution we give a progress report on our systematic study
of a large sample of post-AGB stars. The sample stars were selected on
the basis of their infrared colours and the selection criteria were
tuned to discover objects with hot dust in the system. We started a
very extensive, multi-wavelength programme which includes the analysis
of our radial velocity monitoring; our optical high-resolution
spectra; our groundbased N-band spectral data as well as the Spitzer
full spectral scans; the broad-band SED and the high
spatial-resolution interferometric experiments with the VLTI.  In this
contribution we highlight the main results obtained so far and argue
that all systems in our sample are indeed binaries, which are
surrounded by dusty Keplerian circumbinary discs. The discs play a
lead role in the evolution of the systems.

\keywords{Stars: AGB and post-AGB --
Stars: atmospheres -- 
Stars: evolution 
(Stars:) binaries: spectroscopic 
Techniques: radial velocities
Techniques: interferometric }
}

\maketitle{}

\section{Introduction}
This conference highlighted once more that the processes which govern
the final evolution of low and intermediate mass stars are still poorly understood.
Fundamental uncertainties remain in our understanding of the
internal structure, the (chemical) evolution and the mass-loss processes
of AGB stars. Another well known challenge is the study of PN
formation, where much research is devoted to trying to understand the origin of
the remarkable morphological and kinematical differences between AGB
circumstellar envelopes and their more evolved counterparts. 
During the transition time, the star and circumstellar envelope must
undergo fundamental and rapid changes in structure, mass-loss mode
and geometry which are still badly understood. The debate on which
physical mechanisms are driving the morphology changes gained even
more impetus after the finding that resolved cooler post-AGB stars
or proto-planetary nebulae (PPNe) display a surprisingly wide variety
in shapes and structure, very early in their evolution off the AGB
\citep{sahai03}. 

Impressive kinematic information resulted from the extensive CO survey
of \citet{bujarrabal01}\,: a fundamental property of the
omnipresent fast molecular outflow in PPNe appears to be that it carries a huge
amount of linear momentum, up to a 1000 times the momentum available
for a radiation driven wind.  
Clearly, other momentum sources have to be explored.
Despite intense debate between proponents of binary models 
and advocates of single star models, this issue remains a mystery.
The testing of binary scenarios remains elusive, also because of 
the lack of observational information on binarity in PNe and often
very obscured PPNe.

\begin{figure*}[t!]
\resizebox{6.5cm}{4cm}{\includegraphics[clip=true]{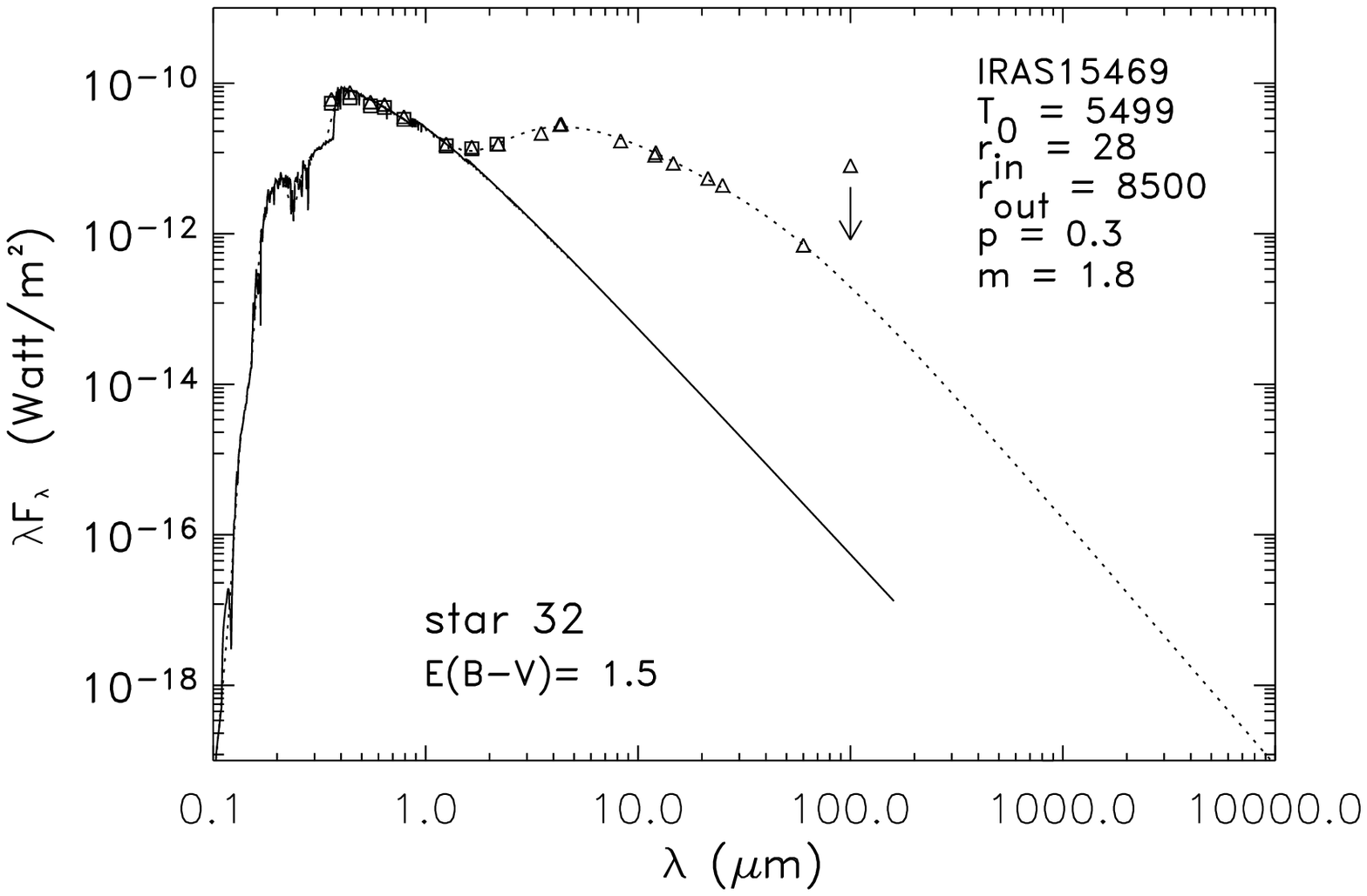}}
\resizebox{6.5cm}{4cm}{\rotatebox{+90}{\includegraphics[clip=true]{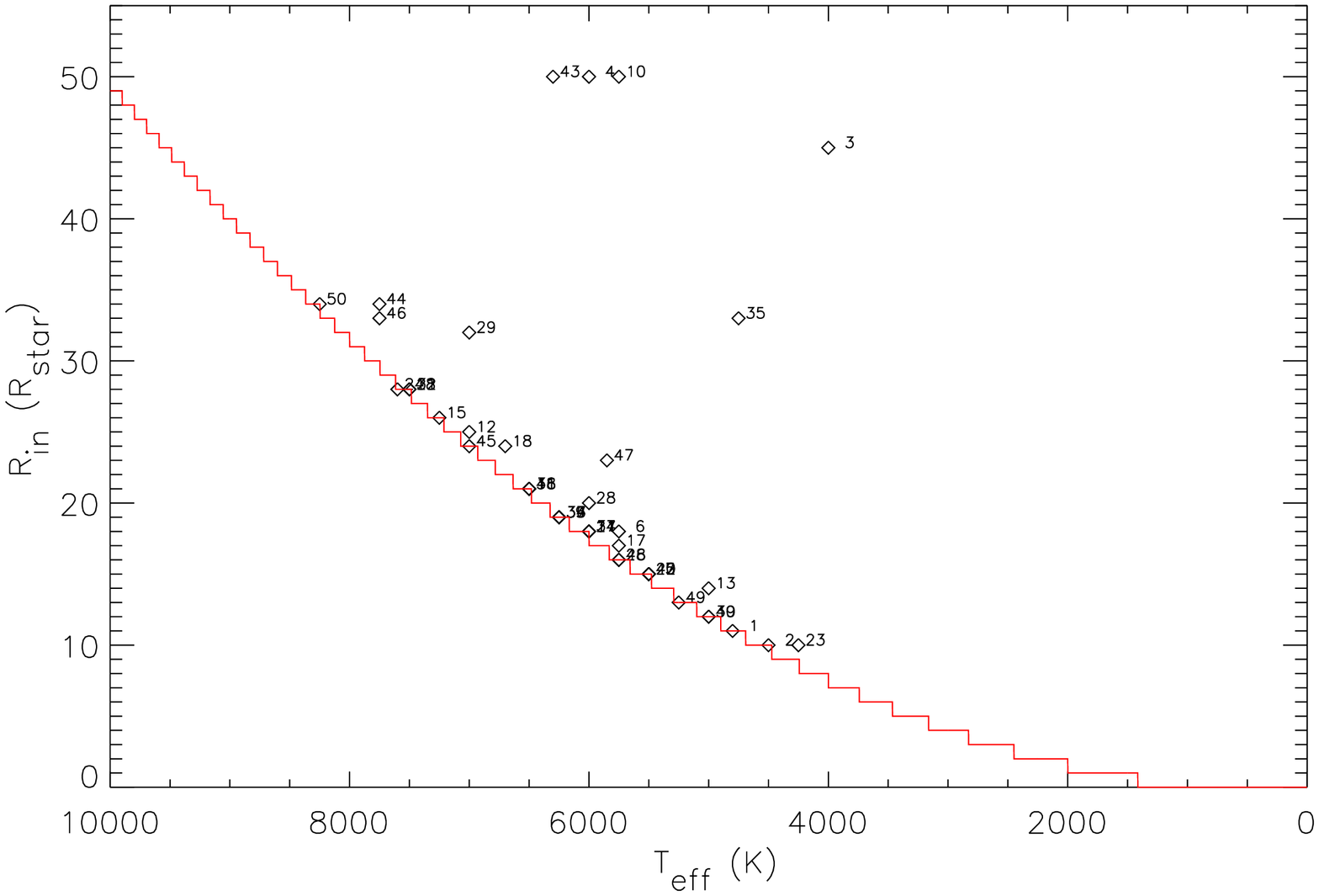}}}
\caption{\footnotesize
The right panel gives an illustrative SED plot of one of the post-AGB
stars (IRAS\,15469-5311). The full line is an appropriate Kurucz model, the measurements
are dereddened. Note the significant dust excess, which starts already
at the H-band. In the right panel we show that for the whole sample
and irrespective of the effective temperature of the central star
(x-axis), the dust excess (R$_{in}$ expressed in stellar radius)
starts  at or very near sublimation temperature (full line) \citep{deruyter06}.}
\label{fig:sed}
\end{figure*}

To study late stellar evolution in binary systems, optically bright,
less obscured post-AGB stars are ideal candidates and in recent years
it became clear that binaries are not uncommon. It was realised that
these binaries have distinct observational characteristics, 
which include broad IR excesses often starting already in H or K,
pointing to the presence 
of both hot and cool dust around the system. It was postulated
that this indicates the presence of gravitationally bound circumstellar material
in the system (for a recent review we refer to \citet{vanwinckel03}).
The most famous example of an evolved binary is the Red Rectangle for which the
Keplerian kinematics of its circumstellar disc have been resolved recently
by interferometric CO measurements \citep{bujarrabal05}.

While the first binary post-AGB stars were serendipitously discovered,
the distinct characteristics of their broad-band SED allowed us to
launch a more systematic search for evolved binaries. In total we selected 51 objects
\citep{deruyter06} which is a fair number compared to some 220
post-AGB  stars known in the Galaxy \citep{szczerba01}.
We started a very extensive multi-wavelength observational study of those systems.
Although the program is far from finished, we give here an overview
our main results obtained so far.

\section{Spectral Energy Distribution (SED)}

The total sample of 51 stars \citep{deruyter06} was defined by uniting
three different sub-groups: the serendipitously discovered binary
post-AGB stars; those RV\,Tauri stars which show an infrared excess as
detected by IRAS; and a new sample of stars \citep{lloydevans99}, detected by IRAS and with
infrared selection criteria tuned to discover new RV\,Tauri stars. 

The RV\,Tauri stars form a rather heterogeneous group of
classical pulsators, with defining light curves showing alternative 
deep and shallow minima. The pulsational periods are between 30 and
150 days. They are generally acknowledged to be evolving on a post-AGB
track, after the detection of thermal infrared radiation of circumstellar
dust around many of them \citep{jura86}. Thanks to the discovery of
RV\,Tauri stars in  the LMC \citep{alcock98}, their location on the
high luminosity end of the Population II Cepheid instability 
strip has been corroborated.

The main result of our systematic study of the broad-band energetics
of the sample stars was that the SEDs of all programme stars are in
fact very similar (see Fig.~\ref{fig:sed}). Double peaked SEDs were not found and 
the dust excess starts at very high temperatures, irrespective of the
effective temperature of the central star. In almost all systems,
dust near sublimation temperature must be present to explain the
very hot dust component. With the expected luminosities and the
effective temperatures of the stars, sublimation temperature is less
than 10 AU from the surface of the star.
Moreover, when available, the long wavelength
fluxes at 850 $\mu$m, are indicative of a component of large, mm-sized
grains \citep{deruyter05}.

In none of the stars there is evidence for a present-day dusty mass
loss; simplified SED modelling showed that in all systems, gravitationally bound
material must be present. The most natural geometry is that the dust
is stored in a Keplerian disc. The large infrared-to-optical
conversion ratios show that inner regions of the disc
must be puffed-up to cover a large enough solid angle, as
seen from the evolved star. In many cases, the large infrared luminosity
is combined with a very small line-of-sight total reddening, which 
gives a useful handle on the likely viewing angle onto the disc.

\section{Radial Velocity Monitoring}

Since the presence of a Keplerian disc is probably a signature of the
binary nature of the central object, we started a long-term radial
velocity monitoring program to detect binary motion and constrain the
orbital parameters. We use a cross-correlation technique with spectral
masks defined by the high signal-to-noise optical spectra obtained
during our chemical analyses program. The photospheric motion of 
pulsationally unstable atmospheres, make the interpretation of radial 
velocity variabitily as due to orbital motion difficult.
Results on individual stars can be found in \citet{maas02,maas03,vanwinckel04}.
Although the program is still running and the full statistics of our
program are not yet determined, we can safely conclude we indeed found
a very high binary rate. For objects with a small pulsational amplitude,
this rate is even 100\% on six stars \citep{vanwinckel06}.

\begin{figure*}[t!]
\resizebox{6.5cm}{5cm}{\rotatebox{+90}{\includegraphics[clip=true]{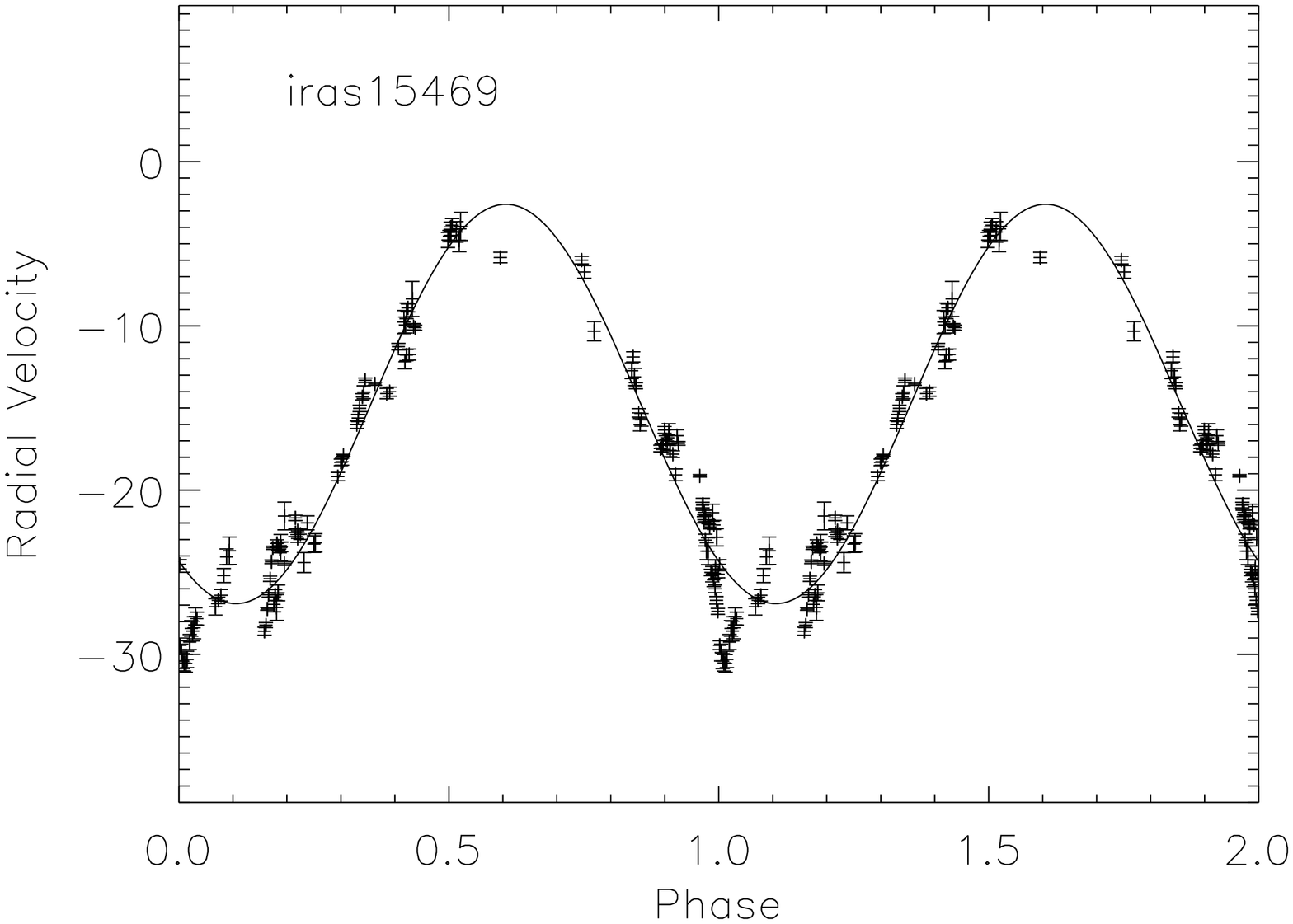}}}
\resizebox{6.5cm}{5cm}{\rotatebox{+90}{\includegraphics[clip=true]{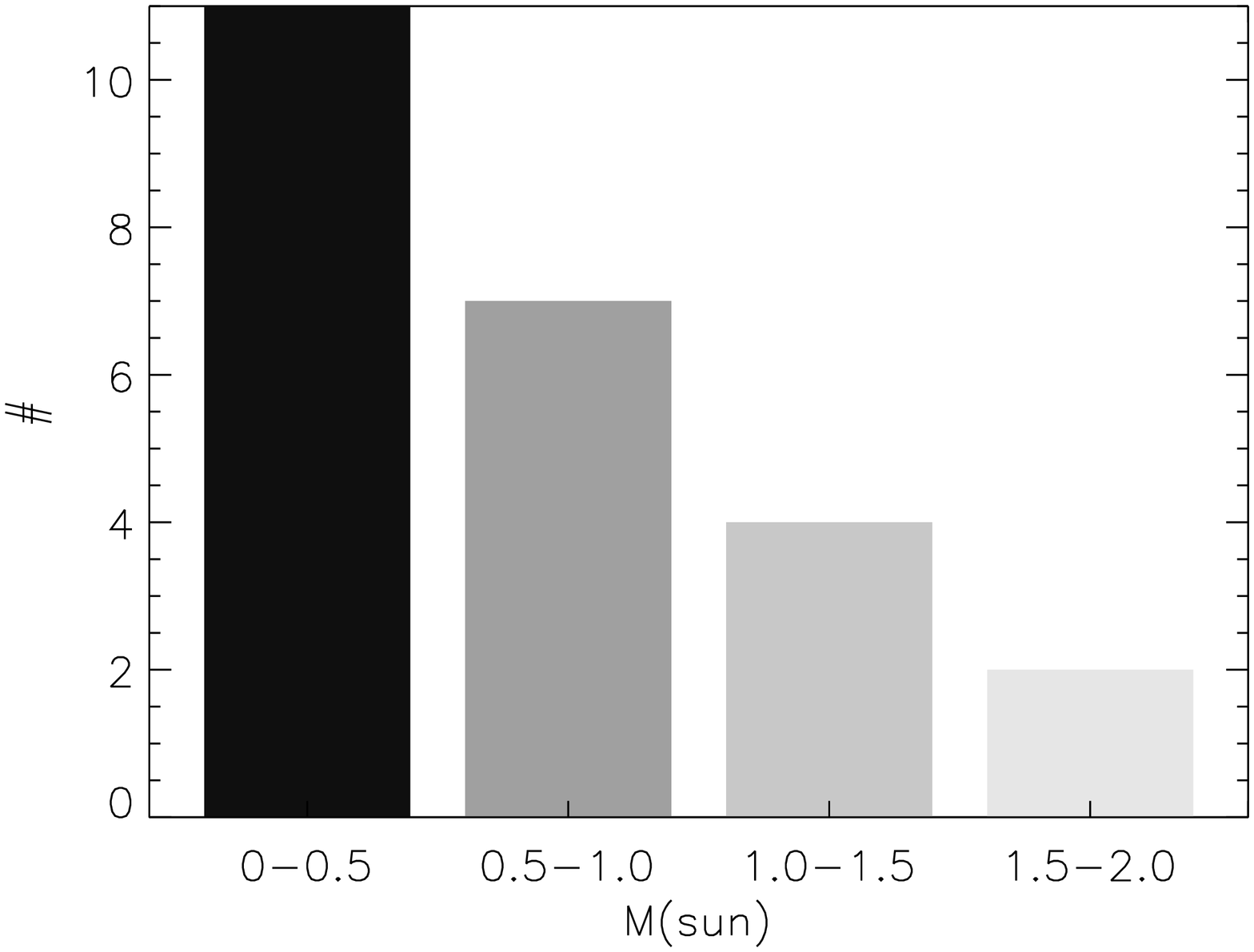}}}
\caption{\footnotesize The left panel shows the orbital solution of
  one of the programme stars (IRAS\,15469-5311). The orbital elements are: Period = 387d
  $\pm$ 1d, amplitude = 12.1$\pm$0.3 km\,s$^{-1}$, system velocity =
  $-$14.8\,$\pm$\,0.2 km\,s$^{-1}$. This star is
  circularised \citep{vanwinckel06}. The right panel shows the
  distribution of systems with respect to the lower
  mass limit of the companion star assuming a 0.6 M$_{\odot}$ primary
  and an inclination of 90$^{\circ}$.}
\label{fig:orbit}
\end{figure*}

In total and complemented with literature values, orbital parameters of
27/51 systems are now determined. The orbital periods are in the range of
100 to 2000 days. In Fig.~\ref{fig:orbit}, we show one of our new
detections. The mass functions give the lower limit of the mass of the
binary companion assuming a 0.6 M$_{\odot}$ primary and an
edge-on orbital plane (Fig.~\ref{fig:orbit}). The masses of the companions cover a
significant range, and in a few systems, the lower mass limit of
the companion is superior to the Chandrasekhar limit of a white dwarf.
In none of the systems is the companion detected or any symbiotic
activity found. Given the orbital characteristics, it is likely that
in most, if not all systems, the companions are non-evolved main-sequence stars.
One of the surprising results is that the orbits are often not
circularised \citep[e.g.][]{vanwinckel03}.

The binaries are now not in contact, but all the orbits are too
short to accommodate a full grown AGB star.
The stars must have been subject to severe interaction in the past
when the primary was at AGB dimensions. It is assumed that during the interaction,
the rather massive circumbinary discs were formed.

\section{Dust Processing}

\begin{figure*}[t!]
\resizebox{6.5cm}{5cm}{\rotatebox{+90}{\includegraphics{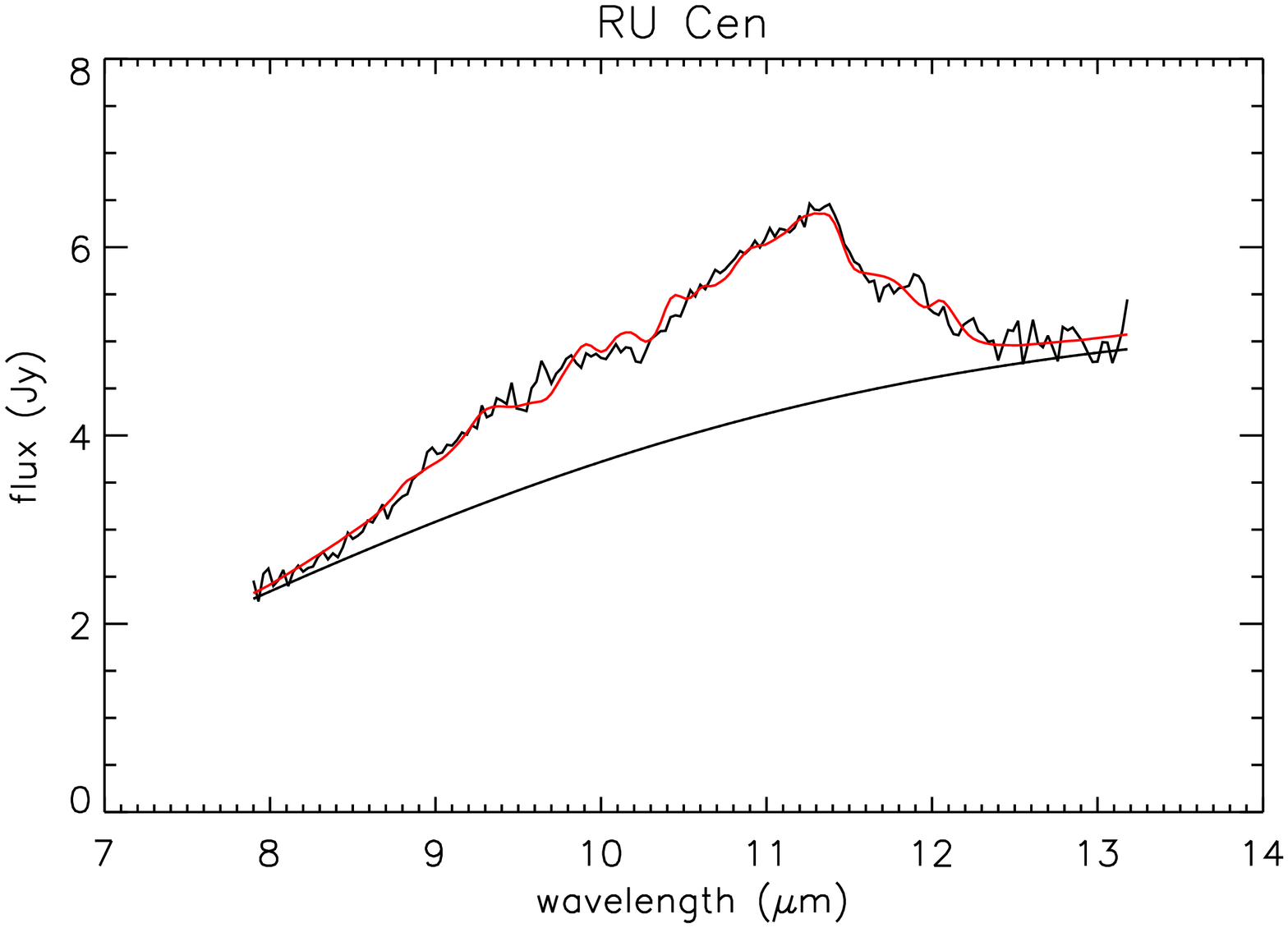}}}
\resizebox{6.5cm}{5cm}{\rotatebox{+90}{\includegraphics{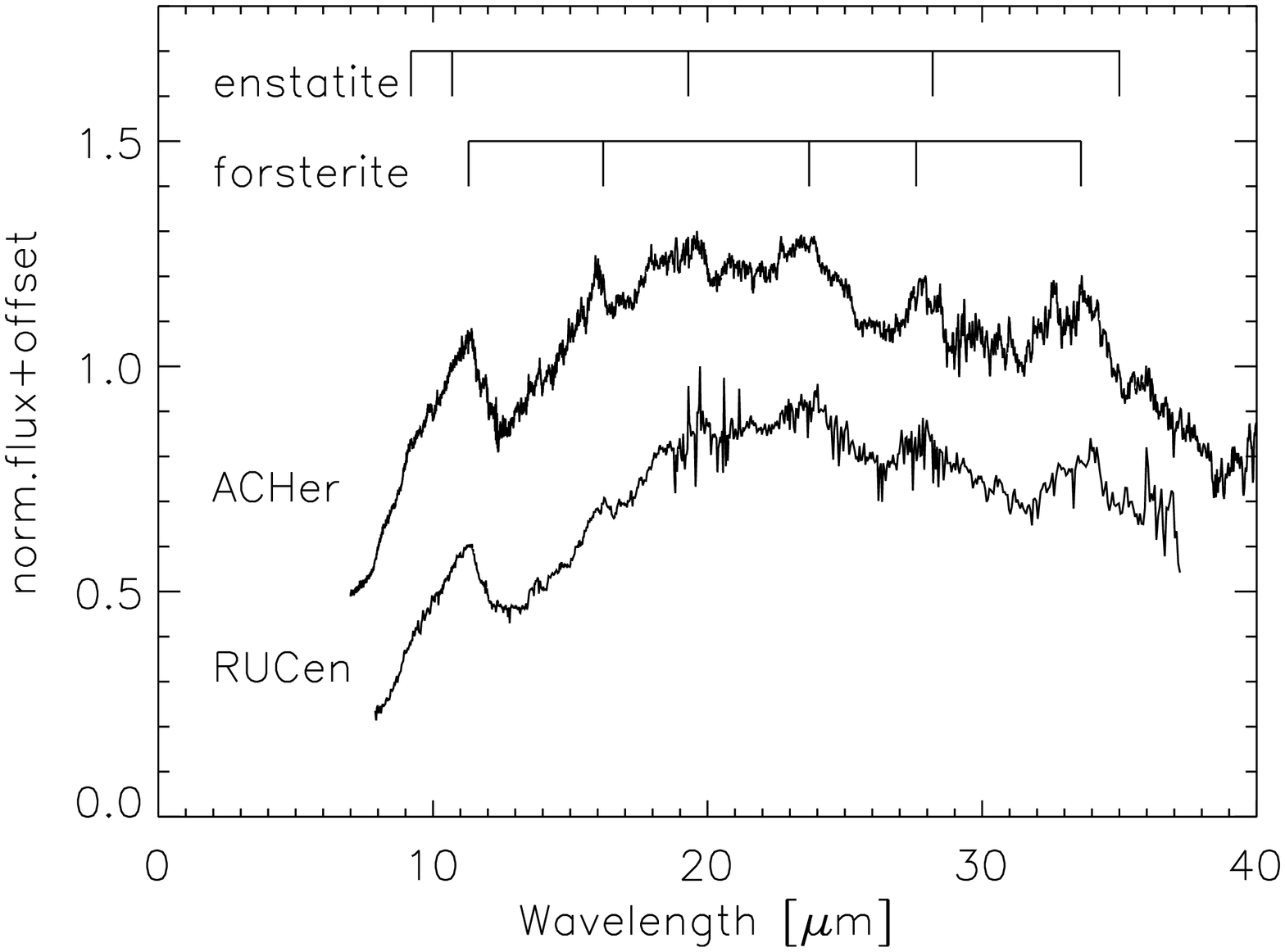}}}
\caption{\footnotesize The N-band spectrum of one of the program stars
(RU\,Cen). The full thick line represents the model in which we used
  several components (amorphous silicates, forsterite, enstatite) and
  different grain sizes. The contributing continuum is also shown \citep{deruyter06b}. 
Our Spitzer high-resolution IRS spectra show  that RU\,Cen is  a spectral analogue
  to another RV\,Tauri star AC\,Her \citep{molster99} which was
  observed by ISO: the silicates
  are very strongly processed both in grain size and certainly in
  crystallinity rate. This again is indicative of processing in gravitationally-bound material.}
\label{fig:timmi}
\end{figure*}

To study the chemico-physical characteristics of
the circumstellar dust, 
the 10 micron N-band is ideally suited, since it
samples resonances of the most dominant silicate minerals (olivines,
pyroxenes and silica) as well as the SiC resonances expected to
prevail in C-rich environments. Moreover, the resonance profiles are very
sensitive to grain sizes from the typical 0.1 $\mu$m up to a few
$\mu$m.  We observed 22 stars from the sample with the TIMMI2
instrument mounted on the 3.6m telescope of ESO. Moreover, a wider
wavelength range was observed for a limited sample during our explorative
Spitzer program.

In all objects studied so far, the dust emission is oxygen rich. The
dust signatures are, however, very different from what is observed in
outflow sources (Fig.~\ref{fig:timmi}).  The narrow resonances of the
crystalline lattices make unique identification possible and
crystalline features are indeed very prominent in the spectra. Moreover, the
width and contrast of the solid state emission profiles are good
tracers of the grain size (e.g. \citet{min04}), while the temperature
structure can be traced by the features' flux ratios.  The main
conclusions of our preliminary analyses of the TIMMI2 and Spitzer
spectra obtained so far (\citet{gielen06, deruyter06b},
Fig.~\ref{fig:timmi}), are that the Mg-rich end members of the
crystalline olivine and pyroxenes (forsterite and enstatite) prevail.
For some sources the silicate emission is over 80\% crystalline !  The
strong crystalline signature and the lack of both amorphous
and crystalline small grains, imply that the grain population is
highly processed \citep{deruyter06b} both in crystallinity and in
average grain size. 

Neither in the recently obtained N-band spectra nor in our Spitzer data, 
are carbon-rich dust or gas signatures found. This despite the
dynamical evidence from the orbits, that at least some objects must
have had initial masses which are theoretically thought to evolve to
carbon stars on the AGB (Fig.~\ref{fig:orbit}). This lack of third
dredge-up enrichment is further evidenced by the CO$_{2}$ gas emission
detected in our Spitzer spectra in one source.  The $^{12}$C/$^{13}$C
is $<$ 10, which illustrates again that $^{12}$C was not enriched
during AGB evolution. Mixed chemistry, which is observed in a few well
studied binaries like HR\,4049 and HD\,44179
\citep[e.g][]{vanwinckel03} seems to be an exception rather than the rule.

Note that our photospheric chemical studies reveal no evidence
for successful 3rd dredge-up enrichment in any of the stars. The study of possible AGB
enrichment is made difficult by the fact that in the total sample, depletion
abundance patterns are frequently detected \citep{vanwinckel03,giridhar05,maas05}. The
basic scenario of the badly understood
depletion process involves a chemical fractionation due to dust formation in
the circumstellar environment followed by a decoupling of the gas and the dust.
The ``cleaned'' gas is then reaccreted on the stellar photosphere, which becomes
depleted of the refractory elements. \citet{waters92} showed that the
most favourable circumstance for this process to occur is, if the circumstellar
dust is trapped in a disc. Possible AGB enrichment could be masked by
this process. S-process elements are refractory, but in all systems of
our sample, the s-process elements show similar abundances to intermediate-mass 
elements with similar condensation temperature.

The photospheric chemical composition as well as the chemical
information of the dust indicate that stellar evolution was shortcut on the
AGB, prior to efficient dredge-up episodes.

\section{Resolving the disc: Interferometry}

Single telescope infrared spectra are of limited diagnostic value to
probe the actual structure and dimension of the circumstellar
environment. The MIDI instrument -- producing spectrally dispersed
fringes in the N-band -- is ideally suited, not only to resolve the
compact discs, but also to obtain spatial information of the different
dust components.  We therefore started an interferometric program on
our sample stars.  With the data available now on only a few objects,
the interferometric measurements prove that the circumstellar emission
originates indeed from a very compact region. \citet{deroo06} showed
that one of the objects, SX\,Cen is not even resolved with a 45m
baseline, implying an upper limit of some 18 A.U. for the diameter of
the dust emission.  Most sample stars are, however, well resolved
giving unprecedented constraints on the spatial dimensions of the
infrared emitting surface.

The added value of the MIDI instrument are the dispersed fringes
allowing us to probe distribution differences between the minerals.
The inner disc reaches temperatures above the glass
temperatures causing the grains to anneal on very short
timescales. Therefore, the innermost disc regions
are expected to be strongly crystalline.
In some objects, this radial gradient in crystallinity is indeed observed 
\citep{deroo06}. Other objects, however, show evidence that the whole
disc is processed since amorphous and crystalline grains show the same
distribution.  The cool crystalline silicates detected in our Spitzer data, as well as
by the interferometric data, show either that radial mixing
was very efficient or that the thermal history of the grains implied
significant processing during the formation process of the grains themselves.

\section{Conclusion}

Although our analysis of all data is not complete yet, it is clear that
all experiments till now confirm that the stars are indeed surrounded by stable
dusty reservoirs in a Keplerian discs.

The global picture that emerges is that a binary star evolved in a
system which is too small to accommodate a full grown AGB star. The
AGB evolution was cut short and during a badly understood phase of
strong interaction, a circumbinary dusty disc was formed, while the
binary system did not suffer dramatic spiral in. What we observe now
is a F-G supergiant in a binary system, which is surrounded by a
circumbinary dusty disc in a bound orbit. In all observed cases, the
dust in the disc is oxygen rich.  Since all systems have orbits well within
the sublimation temperature of dust grains, the dust discs must be
circumbinary.

Formation scenarios of the discs include a wind capture
scenario, in which the dusty wind of the primary is accreted by the
companion, or a scenario in
which the disc is formed through a non-conservative mass transfer in
an interacting binary. Given the orbits detected till now, the second scenario is more
likely, but unfortunately it is theoretically poorly explored.
The thermal history of the grains may have been very
different from that in normal AGB winds, which could lead to very
different chemo-physical properties of the grains during dust
formation.  On the other hand, in the
limited sample for which we have the full Spitzer spectrum, several do 
show evidence of much less processed
material, while the global binary characteristics (orbit,
SED) are not very different. This would suggest also an effect of processing
and radial mixing during the storage time in the disc.

In the limited sample
observed till now, there is no clear connection between the quantified
properties of the dust spectra, and other characteristics of the stars
like the binary orbit, effective temperature of the central star
and/or SED characteristics. Clearly many questions remain.

The formation structure and evolution of the Keplerian disc is
far from being understood, but it does appear to be a key ingredient
in our understanding of the late evolution of a very significant
binary population.

\begin{acknowledgements}
We acknowledge the Mercator telescope team and the many observers of
the 'Instituut voor Sterrenkunde'. Their dedication makes long-term monitoring
programs on binaries with periods of one to several years possible.
We warmly acknowledge all other co-investigators on the versatile
observational programs.
\end{acknowledgements}

\bibliographystyle{aa}

\end{document}